\documentclass[epsfig,12pt]{article}
\usepackage[dvips]{graphicx}
\begin{document} 
 
  
\begin{center}

\LARGE\bf
\rule{0mm}{7mm} Anti-hyperon Polarization 
in $pA$ and $\Sigma^-A$ Collisions and Intrinsic Antidiquark State in Incident Baryon
\end{center}

\begin{center}
Hujio Noda$^1$, Tsutomu Tashiro$^2$ and
Shin-ichi Nakariki$^2$ \\
\vspace{3ex}
$^1$Department of Physics, Ibaraki University, Mito, Ibaraki 310-8512, Japan\\
$^2$Resarch Institute of Natural Sciences, Okayama University of Science, Okayama, Okayama 700-0005, Japan\\
\end{center}
\vspace{2 ex}

\centerline{}


\begin{abstract}
{We discuss the relation between the polarization of inclusively produced (anti-)\\hyperons 
and the incident baryon states in the framework of
the constituent quark-diquark cascade model.
We assume that there is an intrinsic diquark-antidiquark state
in the incident baryon, in which the intrinsic diquark immediately
fragments into a non-leading baryon and the antidiquark behaves as
a valence constituent. It is also assumed that the valence (anti)diquark
in the incident nucleon tends to combine selectively with a spin-down
sea quark and, on the other hand, the spin-up valence quark in the
projectile is chosen by a sea (anti)diquark in preference to the spin-down
valence quark.
It is found that the incident spin-1/2 baryon is mainly composed of
a spin-0 valence diquark and a valence quark, and
contains an intrinsic diquark-antidiquark state with
a probability of about 7$\%$. \\

PACS \\
13.85.Ni(Inclusive production with identified hadrons)\\ 
13.88.+e(Polarization in interactions and scattering)\\
14.20.Jn(Hyperons)
}
\end{abstract}

\vspace{5 ex}
\rule[.5ex]{16cm}{.02cm}
$^1$ email:  noda@mcs.ibaraki.ac.jp \\
$^2$ email:  tashiro@sp.ous.ac.jp \\
$^3$ email:  nakariki@sp.ous.ac.jp \\

%
\setlength{\oddsidemargin}{0 cm}
\setlength{\evensidemargin}{0 cm}
\setlength{\topmargin}{-0.5 cm}
\setlength{\textheight}{22 cm}
\setlength{\textwidth}{16 cm}
\setcounter{totalnumber}{20}
%
\pagestyle{plain}
\setcounter{page}{1}


\newpage
\section{Introduction}
\label{intro}
\indent
It is well known that hyperons produced in unpolarized proton-proton and proton-nucleus ($pA$) collisions are polarized 
transversely to the production plane.  For example, the $\Lambda$ is significantly negatively polarized
and $\bar{\Lambda}$ 
is not polarized\cite{Lesnik,Bunce,Smith,Lundberg,Ramberg94,Heller78}. The polarizations of $\Sigma^\pm$ and 
$\Xi$ produced inclusively in proton beam in the soft hadronic interaction regions are positive 
and negative, respectively\cite{Wilkinson,Deck,Morelos95,Heller83,Rameika86,Duryea91}. 
Anti-hyperons are also polarized. The $\bar{\Sigma}^-$ is polarized positively 
and $\bar{\Xi}^+$ is polarized negatively.\cite{Morelos93,Ho} 
It is important to discuss the anti-hyperon polarization as well as the hyperon polarization. 
On the other hand, single-spin asymmetries of $\pi^\pm$ in $p^\uparrow p$ and $\bar{p}^\uparrow p$ collisions have also 
been observed\cite{Adams91,Adams91-2,Adams04,Bravar}. 
The single-spin asymmetry is defined as $A_N=(\sigma^\uparrow-\sigma^\downarrow)/(\sigma^\uparrow+\sigma^\downarrow)$,
where $\sigma^\uparrow$ ($\sigma^\downarrow$) denotes the cross section of $\pi^+$ 
to go left (right) looking downstream in the $p^\uparrow$ fragmentation region. 
The direction of transverse motion of the produced hadron depends on the polarization of the incident 
hadron. In the case of $p^\uparrow p \rightarrow \pi^+ X$, the produced $\pi^+$ tends to go left looking 
downstream, i.e., $A_N>0$.

Straightforward perturbative QCD (pQCD) and collinear factorization approaches 
underestimate the hyperon polarization in unpolarized $pA$ collisions and single-spin asymmetry \cite{Kane,Dharmaratna_Goldstein90,Dharmaratna_Goldstein96}. 
The single-spin asymmetry was analyzed by the pQCD with the higher twist terms\cite{Qiu_Sterman,Kanazawa_Koike} or with 
the inclusion of spin and transverse-momentum effects in parton distribution\cite{A_B_M_95_99,Anselmino_Boglione_Alesio_Leader_Murgia,Ma_Schmidt_Yang}. 
The pQCD approach with polarizing fragmentation functions was applied to $\Lambda$ and $\bar{\Lambda}$ polarizations 
at $p_T>1$ GeV/c.\cite{Anselmino_Boer_Alesio_Murgia,D'Alesio_Murgia} 
While a number of different approaches have been proposed and applied to the hyperon polarization and the single-spin 
asymmetry\cite{A_G_I,DeGrand_Miettinen,Soffer_Tornqvist,B_P_R,Troshin_Tyurin_P,Boros_Zuo-tang,Zuo-tang_Boros,Hui_Zuo-tang,Yamamoto_Kubo_Toki,Kubo_Yamamoto_Toki,Neal_Burelo}.  
However, they do not explain the data on hyperon and anti-hyperon polarizations satisfactorily. 
In them, C. Boros, L. Zuo-tang and D. Hui pointed out that it is important to consider the
correlation between the spin of the fragmenting valence quark and the direction of momentum of
the outgoing hadron in analyses both of the hyperon polarization in unpolarized $pA$ collisions
and single-spin asymmetry\cite{Boros_Zuo-tang,Zuo-tang_Boros,Hui_Zuo-tang}. Therefore, following this argument, 
in our previous paper \cite{tnnf} we have introduced the spin dependence
in the quark-diquark cascade model.  There we assumed that, when a leading baryon is produced, the valence
diquark in the incident nucleon tends to combine with a spin-down sea quark and the spin-up valence
quark in the projectile is chosen in preference to the spin-down valence
quark by a sea diquark. We successfully analyzed the hyperon polarization, but could not explain
the anti-hyperon polarization.

In this paper, we analyze the anti-hyperon polarization in baryon fragmentation regions by introducing the
intrinsic diquark-antidiquark state in the incident baryon and by using the quark-diquark cascade model with spin. 
We investigate the soft hadronic interactions in the regions of $p_T < 2$GeV/c. We analyze hyperon and   
anti-hyperon polarizations in unpolarized $pA$ and $\Sigma A$ collisions. 
We find that the spin-$\frac{1}{2}$ baryon contains an intrinsic diquark-antidiquark state with a 
probability of about 7$\%$.
In Sec. 2, we describe our model. We assume the existence of the intrinsic antidiquark state in the incident baryon and
introduce spin dependence in the model. In Sec. 3, we analyze data on polarized hyperon and
anti-hyperon productions. Conclusion and discussion are given in Sec. 4.

\section{Model}
\subsection{Structure of the incident baryon}
\begin{figure}
\begin{center}
\resizebox{0.5\textwidth}{!}{%
  \includegraphics{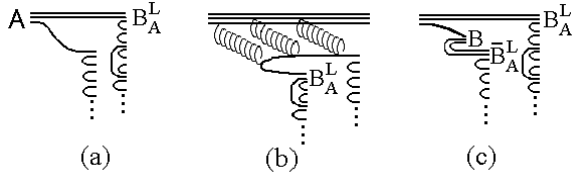}
}
\end{center}
\caption{
Breaking up of incident baryon $A$ into two constituents illustrated by bold lines: 
(a) valence quark and valence diquark, 
(b) quark and antiquark, 
(c) intrinsic diquark and intrinsic antidiquark with a non-leading baryon $B$.
 $B^L_A$ 
denotes the leading baryon in the fragmentation region of $A$.
}
\label{fgr:prj_brk}       
\end{figure}

We treat the incident baryon as to be 
the superposition of a valence quark-diquark 
and a valence quark-diquark with an intrinsic diquark-antidiquark state: 
\begin{eqnarray}
|B>_{in}=c_0|(qq)_Vq_V>
+c_1|(qq)_Vq_V(qq)(\overline{qq})>. 
\label{eqn:intr_qq}
\end{eqnarray}
The probabilities of the intrinsic diquark-antidiquark state in the incident baryon is denoted by $c_1^2(1-P_{PM})$. 
Here $P_{PM}$ denotes the probability for the Pomeron exchange process.
It is assumed that
the intrinsic diquark immediately combines with the valence quark and converts into 
a non-leading baryon, and the remaining intrinsic antiquark (antidiquark)
behaves as a valence constituent. In order to explain the anti-hyperon polarization, the remaining intrinsic antidiquark 
is assumed to be in a spin-1 state. 
Let us consider the hyperon productions in the inclusive reaction $ A+B \rightarrow C+X$ in the 
beam fragmentation region at $p_T < 2$GeV/c. 
When a collision occurs, the incident baryon breaks up into two constituents, as shown in Fig.\ref{fgr:prj_brk}: 
(a) breaking into $(qq)_V$ and $q_V$ with probability $(1-c_1^2)(1-P_{PM})$, 
(b) emission of gluons converting into a quark-antiquark pair with probability $P_{PM}$, 
and (c) emission of a non-leading baryon and breaking into 
a valence diquark and an intrinsic antidiquark with probability $c_1^2(1-P_{PM})$. 

\subsection{Introduction of spin into the incident baryon and the cascade process}
We extended the quark-diquark cascade model\cite{tnki} 
to include the spin dependence by using
SU(6) wave functions for hadrons and the quark-diquark representation for baryons as discussed in Ref.\cite{Fisjak_Kistenev}. 
 We then applied our model to hyperon polarization in $hA$ collisions.\cite{tnnf} 
The wave function for the incident $p^\uparrow$ in the state represented by the first term of (\ref{eqn:intr_qq}) is written as
\begin{eqnarray}
|p^\uparrow>
&=&\sqrt{\frac{2}{9}}\sin\theta\big(\sqrt2|\{uu\}^V_{11}d_V^\downarrow> - |\{uu\}^V_{10}d_V^\uparrow> 
- |\{ud\}^V_{11}u_V^\downarrow> + \sqrt{\frac{1}{2}}|\{ud\}^V_{10}u_V^\uparrow>\big) \nonumber \\
& +& \cos\theta|[ud]^V_{00}u_V^\uparrow>, 
\label{eqn:pup_su6}
\end{eqnarray}
where the parameter $\theta$ is the mixing angle.
The incident spin-up proton ($p^\uparrow$) breaks up into a valence quark and a valence diquark: 
\begin{eqnarray}
  d_V^\downarrow+\{uu\}^V_{11}, d_V^\uparrow+\{uu\}^V_{10}, u_V^\downarrow+\{ud\}^V_{11},
  u_V^\uparrow+\{ud\}^V_{10},  u_V^\uparrow+[ud]^V_{00}, \nonumber
\label{eqn:p_up2dqdq}
\end{eqnarray}
with probabilities $(1-c_1^2)(1-P_{PM})$ times 
$\frac{4}{9}\sin^2\theta, \frac{2}{9} \sin^2\theta, \frac{2}{9}\sin^2\theta, \frac{1}{9}\sin^2\theta$ and 
$\cos^2\theta$, 
respectively and into a quark and an antiquark:
\begin{eqnarray}
 u^\uparrow+\bar{u}^\downarrow,~u^\downarrow+\bar{u}^\uparrow,~d^\uparrow+\bar{d}^\downarrow,~
 d^\downarrow+\bar{d}^\uparrow,~s^\uparrow+\bar{s}^\downarrow,~s^\downarrow+\bar{s}^\uparrow   \nonumber
\end{eqnarray}
with probabilities $P_{PM}/2$ times $P_{u\bar{u}},P_{u\bar{u}},P_{d\bar{d}}, P_{d\bar{d}},P_{s\bar{s}}$ 
and $P_{s\bar{s}}$, respectively. 
It also breaks up into a valence diquark, a intrinsic antidiquark and a non-leading baryon 
with probabilities $c_1^2(1-P_{PM})$ times the corresponding diquark-antidiquark pair creation probabilities.
The brackets $[~]$ and $\{\}$ for the diquark states denote flavor
anti-symmetric and symmetric states, respectively, and subscripts denote the spin states.

Hadrons are produced by the cascade processes as follows; \\
(i) baryon productions
\noindent
\begin{eqnarray}
q^\uparrow & \rightarrow &  B_{\frac{1}{2}\frac{1}{2}}(q[q'q''])+[\overline{q'q''}]_{00}, B_{\frac{1}{2}\frac{-1}{2}}(q\{q'q''\})+\{\overline{q'q''}\}_{11}, \nonumber \\  
           &            &  B_{\frac{3}{2}\frac{3}{2}}(q[q'q''])+\{\overline{q'q''}\}_{1-1}, B_{\frac{3}{2}\frac{1}{2}}(q[q'q''])+\{\overline{q'q''}\}_{10},~~~ \nonumber \\  
           &            &  B_{\frac{3}{2}\frac{-1}{2}}(q[q'q''])+\{\overline{q'q''}\}_{11},  
\label{eqn:q2B}
\end{eqnarray}
\begin{eqnarray}
{[q'q'']}_{00}& \rightarrow & B^\uparrow(q[q'q''])+\bar{q}^\downarrow, B^\downarrow(q[q'q''])+\bar{q}^\uparrow,~~~~~~~~~~~~~~~~~~~~~~~~ \nonumber \\
\{q'q''\}_{11}& \rightarrow & B_{\frac{3}{2}\frac{3}{2}}(q\{q'q''\})+\bar{q}^\downarrow, B_{\frac{3}{2}\frac{1}{2}}(q\{q'q''\})+\bar{q}^\uparrow, \nonumber \\ 
&&\cdot \cdot \cdot, 
\label{eqn:dq2B}
\end{eqnarray}
(ii) meson productions 
\begin{eqnarray}
q^\uparrow & \rightarrow & M_{00}(q\bar{q})+q'^\uparrow,  M_{10}(q \bar{q}')+q'^\uparrow, M_{11}(q \bar{q}')+q'^\downarrow, ~~~~~~~~~~~~~~\nonumber \\
          &            &  M_{20}(q \bar{q}')+q'^\uparrow, M_{21}(q \bar{q}')+q'^\downarrow, 
\label{eqn:q2M}
\end{eqnarray}
\begin{eqnarray}
{[q'q'']}_{00} & \rightarrow &  M_{00}(q\bar{q}')+[qq'']_{00}, M_{10}(q\bar{q}')+\{qq''\}_{10}, ~~~~~~~~~~~~~~\nonumber \\
         &           &  M_{11}(q\bar{q}')+\{qq''\}_{1-1}, \nonumber  \\ 
\{q'q''\}_{11} & \rightarrow &  M_{11}(q\bar{q}')+[qq'']_{00} ,M_{11}(q\bar{q}')+\{qq''\}_{10}, ~~~~~~~~~~~~\nonumber \\
         &           &  M_{10}(q\bar{q}')+\{qq''\}_{11}, M_{21}(q\bar{q}')+\{qq''\}_{10},\nonumber\\
&& \cdot \cdot \cdot ,
\label{eqn:dq2M}
\end{eqnarray}
where $q$ denotes $u, d$ and $s$ and $[q'q'']$ does $[ud], [us]$ and 
$[ds]$ and so on. 
$ \epsilon$ 
and $ 1-\eta $ denote probabilities of baryon production
from a quark and a diquark, respectively. 
We consider only j=3/2 decuplet and j=1/2 octet baryons as produced baryons.

From the isospin invariance,
the $q\bar{q}$ pair creation 
probabilities are $  P_{u\bar{u}}=P_{d\bar{d}}$ and $P_{s\bar{s}}=1-2P_{u\bar{u}}$. 
The 
probabilities of $ [qq']\overline{[qq']}, \{qq'\}\overline{\{qq'\}} $ 
and $ \{qq\}\overline{\{qq\}} $ pair creations from a quark are 
chosen as $ \epsilon P_{q\bar{q}}P_{q'\bar{q}'}, \epsilon P_{q\bar{q}}P_{q'\bar{q}'}$ 
and $\epsilon {P_{q\bar{q}}}^2$, respectively.  
For example, baryons are produced from $[ud]_{00}$ and $\{uu\}_{11}$ as follows:
\begin{eqnarray}
[ud]_{00} \rightarrow p^\uparrow+\bar{u}^\downarrow, p^\downarrow+\bar{u}^\uparrow, n^\uparrow+\bar{d}^\downarrow, n^\downarrow+\bar{d}^\uparrow, 
\Lambda^\uparrow+\bar{s}^\downarrow, \Lambda^\downarrow+\bar{s}^\uparrow,
\label{eqn:ud00}
\end{eqnarray}
with probabilities 
\begin{eqnarray}
N_1 P_{u\bar{u}}/2, N_1 P_{u\bar{u}}/2, N_1 P_{d\bar{d}}/2, N_1 P_{d\bar{d}}/2,
N_1 P_{s\bar{s}}/3, N_1 P_{s\bar{s}}/3, 
\label{eqn:Pud00}
\end{eqnarray}
respectively, where $
N_1=(1-\eta)/(P_{u\bar{u}}+P_{d\bar{d}}+\frac{2}{3}P_{s\bar{s}})$, 
and
\begin{eqnarray}
&\{uu\}_{11} & \rightarrow p^\uparrow+\bar{d}^\uparrow, \Sigma^{+\uparrow}+\bar{s}^\uparrow, \Delta^{++}_{\frac{3}{2}\frac{3}{2}}+\bar{u}^\downarrow, \Delta^{++}_{\frac{3}{2}\frac{1}{2}}+\bar{u}^\uparrow, \nonumber \\
&&  \Delta^{+}_{\frac{3}{2}\frac{3}{2}}+\bar{d}^\downarrow, \Delta^{+}_{\frac{3}{2}\frac{1}{2}}+\bar{d}^\uparrow, \Sigma^{*+}_{\frac{3}{2}\frac{3}{2}}+\bar{s}^\downarrow, \Sigma^{*+}_{\frac{3}{2}\frac{1}{2}}+\bar{s}^\uparrow, 
\label{eqn:uu11}
\end{eqnarray}
with probabilities 
\begin{eqnarray}
&&2N_2 P_{d\bar{d}}/9, 2N_2 P_{s\bar{s}}/9, N_2 P_{u\bar{u}}, N_2 P_{u\bar{u}}/3, \nonumber \\
&&N_2 P_{d\bar{d}}/3, N_2 P_{d\bar{d}}/9, N_2 P_{s\bar{s}}/3, N_2 P_{s\bar{s}}/9,
\label{eqn:Puu11}
\end{eqnarray}
respectively, where $
N_2=(1-\eta)/(\frac{4}{3}P_{u\bar{u}}+\frac{2}{3}P_{d\bar{d}}+\frac{2}{3}P_{s\bar{s}})$. Similarly, from $s^\uparrow$, we have
\begin{eqnarray}
& s^\uparrow & \rightarrow \Lambda^\uparrow+[\overline{ud}]_{00}, \Sigma^{+\uparrow}+\{\overline{uu}\}_{10}, \Sigma^{+\downarrow}+\{\overline{uu}\}_{11}, \nonumber \\
&&\cdot\cdot\cdot,\nonumber \\
&&\Xi^{*-}_{\frac{3}{2}\frac{3}{2}}+\{\overline{ds}\}_{1-1}, \Xi^{*-}_{\frac{3}{2}\frac{1}{2}}+\{\overline{ds}\}_{10}, \Xi^{*-}_{\frac{3}{2}\frac{-1}{2}}+\{\overline{ds}\}_{11}, \nonumber \\
&&\Omega^{-}_{\frac{3}{2}\frac{3}{2}}+\{\overline{ss}\}_{1-1}, \Omega^{-}_{\frac{3}{2}\frac{1}{2}}+\{\overline{ss}\}_{10}, \Omega^{-}_{\frac{3}{2}\frac{-1}{2}}+\{\overline{ss}\}_{11}, 
\label{eqn:s_up}
\end{eqnarray}
with probabilities 
\begin{eqnarray}
&&\frac{1}{3}N_3 P_{u\bar{u}}P_{d\bar{d}}, \frac{1}{9}N_3 {P_{u\bar{u}}}^2, \frac{2}{9}N_3 {P_{u\bar{u}}}^2, \nonumber \\
&&\cdot\cdot\cdot,\nonumber \\
&&\frac{2}{3}N_3 P_{d\bar{d}}P_{s\bar{s}}, \frac{4}{9}N_3 P_{d\bar{d}}P_{s\bar{s}}, \frac{2}{9}N_3 P_{d\bar{d}}P_{s\bar{s}}, \nonumber \\
&&N_3 {P_{s\bar{s}}}^2, \frac{2}{3}N_3 {P_{s\bar{s}}}^2, \frac{1}{3}N_3 {P_{s\bar{s}}}^2, 
\label{eqn:Ps_up}
\end{eqnarray}
respectively. Here, $N_3=\epsilon/(\frac{5}{3}({P_{u\bar{u}}}^2+{P_{d\bar{d}}}^2)+
\frac{4}{3}P_{u\bar{u}}P_{d\bar{d}}+2(P_{u\bar{u}}+P_{d\bar{d}}+P_{s\bar{s}})P_{s\bar{s}})$.

In the final step of the cascade process, we assume that the constituents recombine into hadrons 
by the following processes: 
\begin{eqnarray}
     q^\uparrow + \bar{q}'^\uparrow &\rightarrow& M_{11}(q\bar{q}'), M_{21}(q\bar{q}'),~\nonumber \\ 
     q^\uparrow + [q'q'']_{00} &\rightarrow& B^\uparrow(q[q'q'']_{00}), ~\nonumber \\
     q^\uparrow + \{q'q''\}_{10} &\rightarrow& B^\uparrow(q\{q'q''\}_{10}), B_{\frac{3}{2}\frac{1}{2}}(q\{q'q''\}_{10}),~\nonumber \\
     \{qq'\}_{10} + [\overline{q''q'''}]_{00} &\rightarrow& B^\uparrow(q''''\{q'q''\}) + \bar{B}^\downarrow(\bar{q}''''[\overline{q''q'''}]) 
\label{eqn:rcmb}
\end{eqnarray}
and so on.  For example, the constituents $d^\downarrow$ and $\{us\}_{11}$ recombine as 
\begin{eqnarray}
 d^\downarrow + \{us\}_{11} &\rightarrow& \Sigma_{\frac{3}{2}\frac{1}{2}}^{*0}, \Sigma_{\frac{1}{2}\frac{1}{2}}^0, \Lambda_{\frac{1}{2}\frac{1}{2}}^0  
\end{eqnarray}    
with probabilities $\frac{1}{2}, \frac{1}{8}$ and $\frac{3}{8}$, respectively. 
The momenta of the recombined hadrons are the sum of those of the final 
constituents. In order to put the recombined hadrons
on the mass shell, we multiply the momenta of all produced hadrons by a common factor 
so that the summation of energies of the produced hadrons 
is equal to the center of mass energy $ \sqrt{s} $. 
Resonances directly produced by the above processes (\ref{eqn:q2B})-(\ref{eqn:dq2M}) and (\ref{eqn:rcmb}) decay into stable particles. 
Here, we note that we use SU(6) wave functions for calculating baryon production ratios, i.e., $\cos^2\theta=\frac{1}{2}$.

The normalized distribution functions of the constituents in the incident baryon, 
which are composed of $ q_V $ and $ (q'q'')^V$, are expressed in terms of the light-like fraction $z$ as
\begin{eqnarray}
 n_{q_V/B}(z) = n_{(q'q'')^V/B}(1-z) = \frac{z^{\beta_{q_V}-1}(1-z)^{\beta_{(q'q'')^V}-1}}{B(\beta_{q_V},\beta_{(q'q'')^V})}. 
\label{eqn:n_{q/B}}
\end{eqnarray}
The dynamical parameters $ \beta$'s are 
related to the intercepts of the Regge trajectories as 
$\beta_u=\beta_d=1-\alpha_{\rho-\omega}(0)\approx 0.5$ and $\beta_s=1-\alpha_\phi(0)\approx1.0$.\cite{chkp,minakata} 
We put 
$\beta_{[q'q'']^V}=1.5(\beta_q'+\beta_{q''})$ and 
$\beta_{\{q'q''\}^V}=2.0(\beta_q'+\beta_{q''})$ for anti-symmetric and symmetric valence diquarks, respectively. 
Similarly, the momentum sharing function of $j$ for the cascade proces $ i \rightarrow H(i\bar{j}) + j$
is assumed as
\begin{eqnarray}
 F_{j/i}(z) = \frac{z^{\gamma\beta_j-1}(1-z)^{\beta_j+\beta_j-1}}{B(\gamma\beta_i,\beta_i+\beta_j)}, 
\label{eqn:Fqq} 
\end{eqnarray}
where $\gamma$ is chosen to be 1.75. 
The factors 1.5, 2.0 and 1.75 are extracted from 
the analysis of hadron spectra in $p$ fragmentation region.\cite{tnki} 
The transverse momentum distribution of the hadron $H$ in the cascade processes is given by the distribution function 
\begin{eqnarray}
   G(\mbox{\boldmath$p$}_{T}^2)=\frac{\sqrt{m_H}}{\alpha}\exp(-\frac{\alpha}{\sqrt{m_H}}\mbox{\boldmath$p$}_{T}^2)
\label{eqn:pT2} 
\end{eqnarray}
in $p_T^2$ space where $m_H$ denotes the mass of $H$. The parameter 
$\alpha$ is fixed to $\alpha=1.8$ GeV$^{-\frac{3}{2}}$ by using experimental data on ${\mbox{\boldmath$p$}_{T}^2}$ 
distributions of pions in $\pi p$ collisions.\cite{tnnik}  
Details of energy-momentum distributions and sharing used in 
our model have been provided in Refs.\cite{tnki,tnnik}. 

\subsection{Quantization direction and spin asymmetry constraint}
The quantization direction for the polarization of an 
observed particle $C$ is defined as the direction $\mbox{\boldmath{$\hat{n}$}}=\mbox{\boldmath{$p$}}_{inc}\times\mbox{\boldmath{$p$}}_{out}/
|\mbox{\boldmath{$p$}}_{inc}\times\mbox{\boldmath{$p$}}_{out}|$, where 
$\mbox{\boldmath{$p$}}_{inc}$ and $\mbox{\boldmath{$p$}}_{out}$ are the momenta of the projectile $A$ ($B$) and 
the produced hadron $C$ in $A$ ($B$) fragmentation region, respectively. 
The collision axis is chosen to be parallel to the $z$-axis and the azimuthal angle of $\mbox{\boldmath{$\hat{n}$}}$ is denoted by $\varphi$. 
We use the sign convention in which a positive polarization is in the same direction as $\mbox{\boldmath{$\hat{n}$}}$.

The baryon (anti-baryon) composed of a diquark (antidiquark), which is converted from the valence diquark 
(antidiquark) only through the cascade processes (\ref{eqn:dq2M}), 
is also treated as a leading baryon (anti-baryon). 
We assume that the quantization direction for the reaction is the normal of 
the production plane of one of the leading hadrons with non zero transverse momentum from $A$ and $B$. 
The leading hadron on the most massive cascade chain is selected 
to determine the spin quantization axis.
When two leading hadrons are produced in the most massive cascade chain, the one 
with the larger energy is selected. 
The quantization direction of the reaction is chosen as the normal of the production plane of the selected leading hadron 
$\mbox{\boldmath{$\hat{n}$}}_L=\mbox{\boldmath{$p$}}_{inc}\times\mbox{\boldmath{$p$}}_{L}
/|\mbox{\boldmath{$p$}}_{inc}\times\mbox{\boldmath{$p$}}_{L}|$. 
Here $\mbox{\boldmath{$p$}}_L$ is the momentum of the selected leading hadron $h^A_L$ ($h^B_L$) and 
$\mbox{\boldmath{$p$}}_{inc}$ is the momentum of the projectile $A$ ($B$) for the selected leading hadron 
$h^A_L$ ($h^B_L$) in $A$ ($B$) fragmentation region. 
The azimuthal angle of $\mbox{\boldmath{$\hat{n}$}}_{L}$ is denoted by $\varphi_L$.

The spin states of the produced hadrons with respect to the direction $\mbox{\boldmath{$\hat{n}$}}_{L}$ are 
determined from the occurrence probabilities of the processes (\ref{eqn:q2B})-
(\ref{eqn:dq2M}) and (\ref{eqn:rcmb}). 
The observed polarization is the difference between the cross section of 
the baryon (anti-baryon) with an upward spin and that of the baryon with a downward spin with respect to the direction 
of its production plane.
In our calculation, we first determine spin 
states with respect to $\mbox{\boldmath{$\hat{n}$}}_{L}$ and then flip the spin of $C$ 
according to the difference between azimuthal angles 
$|\Delta \varphi|=|\varphi-\varphi_L|$, in order to obtain the spin state with respect 
the direction $\mbox{\boldmath{$\hat{n}$}}$. 

\begin{figure}
\begin{center}
\resizebox{0.5 \textwidth}{!}{%
  \includegraphics{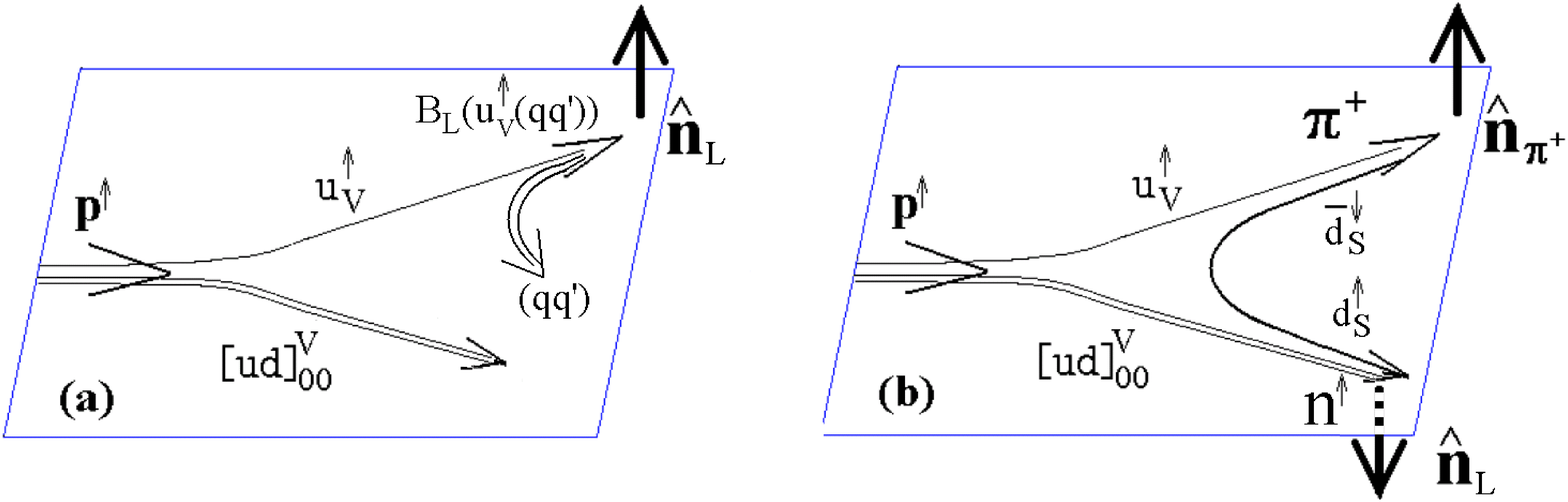}
}
\end{center}
\caption{
Production planes of leading particles in the fragmentation region of $p^\uparrow$; (a)  Production of the leading baryon 
$B_L(u_V^\uparrow(qq'))$ from the valence quark $u_V^\uparrow$.
(b) The exclusive process $p^\uparrow p \rightarrow \pi^+ n^\uparrow p$. 
The valence diquark $[ud]^V_{00}$ picks up a spin-down sea quark $d_{S}$
with respect to the direction ${\mbox{\boldmath{$\hat{n}_L$}}}$. 
}
\label{fgr:dq_q2B}       
\end{figure}

The hyperon polarization 
in an unpolarized $pA$ 
collision is related to the single-spin 
asymmetry because of the correlation
between the spin of the fragmenting valence quark 
and the direction of the momentum of the outgoing hadron\cite{Boros_Zuo-tang,Zuo-tang_Boros,Hui_Zuo-tang}.
The positive value of the single-spin asymmetry of $\pi^+$ in  $p^\uparrow p $ collisions implies that the valence $u_V^\uparrow$ 
quark tends to go left and the valence diquark $[ud]^V_{00}$ tends to go right looking downstream, as shown in Fig.\ref{fgr:dq_q2B}(a), 
provided the transverse momentum is conserved during the breaking up of the projectile $p^\uparrow$.
When the valence $u_V^\uparrow$ quark and a sea diquark $(qq')$ produce a leading baryon $B_L(u_V^\uparrow(qq'))$ in the direction 
of $u_V^\uparrow$, the valence quark tends to be in the spin-up state with respect to the direction 
$\mbox{\boldmath{$\hat{n}$}}_L=\mbox{\boldmath{$p$}}_{inc}\times\mbox{\boldmath{$p$}}_{L}
/|\mbox{\boldmath{$p$}}_{inc}\times\mbox{\boldmath{$p$}}_{L}|$, i.e., the direction of the normal of the 
production plane of $B_L(u_V^\uparrow(qq'))$, as shown in Fig.\ref{fgr:dq_q2B}(a).
Then, we assume that the spin-up valence quark in the incident hadron is chosen
in preference to the spin-down valence quark by a sea diquark to form a leading
baryon, with respect to the direction $\mbox{\boldmath{$\hat{n}_L$}}$.
When we consider the exclusive process $p^\uparrow \rightarrow \pi^+ + n^\uparrow$ and neglect the angular 
momentum effect, $\pi^+$ tends to go left 
and $n^\uparrow$ tends to go right looking downstream, as shown in Fig.\ref{fgr:dq_q2B}(b). 
According to this picture, 
we assume that 
the valence diquark in the incident proton tends to pick up a spin-down sea quark to form a leading baryon 
with respect 
to the direction $\mbox{\boldmath{$\hat{n}$}}_L$.

The probabilities of symmetric and anti-symmetric valence diquarks picking up a sea $s^\uparrow$
quark forming a leading baryon are 
given by $P_{\{\}^V}^{s^\uparrow}=(1+C_{\{\}}^s)/2$ and $P_{[~]^V}^{s^\uparrow}=(1+C_{[~]}^s)/2$, respectively. 
The probabilities of symmetric and anti-symmetric valence diquarks picking up $u^\uparrow$ or $d^\uparrow$ are 
given by $P_{\{\}^V}^{q^\uparrow}=(1+C_{\{\}}^q)/2$ and $P_{[~]^V}^{q^\uparrow}=(1+C_{[~]}^q)/2$, respectively. 
Hereafter, $q$ denotes $u$ or $d$ quark.
 For the cases of the valence diquarks $[ud]^V_{00}$ in (\ref{eqn:ud00}) and $\{uu\}^V_{11}$ in (\ref{eqn:uu11}) 
 forming a leading baryon with a sea quark, 
the probabilities (\ref{eqn:Pud00}) and (\ref{eqn:Puu11}) are changed to
\begin{eqnarray}
&&\frac{N_1}{2} P_{u\bar{u}}(1+C_{[~]}^q), \frac{N_1}{2} P_{u\bar{u}}(1-C_{[~]}^q), 
\frac{N_1}{2} P_{d\bar{d}}(1+C_{[~]}^q), \frac{N_1}{2} P_{d\bar{d}}(1-C_{[~]}^q), \nonumber \\
&&\frac{N_1}{3} P_{s\bar{s}}(1+C_{[~]}^s), \frac{N_1}{3} P_{s\bar{s}}(1-C_{[~]}^s), 
\label{eqn:Pud00a}
\end{eqnarray}
and
\begin{eqnarray}
&&\frac{2}{9}N'_2 P_{u\bar{u}}(1+C_{\{\}}^q), \frac{2}{9}N'_2 P_{s\bar{s}}(1-C_{\{\}}^s), \nonumber \\
&&N'_2 P_{u\bar{u}}(1+8C_{\{\}}^q), \frac{1}{3}N'_2 P_{u\bar{u}}(1-C_{\{\}}^q), \frac{1}{3}N'_2 P_{d\bar{d}}(1+C_{\{\}}^q), \nonumber \\
&&\frac{1}{9}N'_2 P_{d\bar{d}}(1-C_{\{\}}^q), \frac{1}{3}N'_2 P_{s\bar{s}}(1+C_{\{\}}^s), \frac{1}{9}N'_2 P_{s\bar{s}}(1-C_{\{\}}^s), 
\label{eqn:Puu11s}
\end{eqnarray}
respectively.

When the selected leading baryon is produced from the valence quark in the incident baryon
and a sea diquark $(ij)$, the probabilities of
the spin state of the valence quark being up and down are given by
$P_{q_V^\uparrow}^{(ij)}=(1+C_q^{(ij)})/2$ and $(1-C_q^{(ij)})/2$ for valence $u$ and $d$ quarks, and
$P_{s_V^\uparrow}^{(ij)}=(1+C_s^{(ij)})/2$ and $(1-C_s^{(ij)})/2$ for a valence $s$ quark,
respectively. That is, the probabilities of the spin-up valence quark $q$ ($s$) going left and right
are $P_{q_V^\uparrow}^{(ij)}=(1+C_q^{(ij)})/2$ and $(1-C_q^{(ij)})/2$
 ($P_{s_V^\uparrow}^{(ij)}=(1+C_s^{(ij)})/2$ and $1-P_{s_V^\uparrow}^{(ij)}$), respectively.
For simplicity, we assume the relations between $P_{q_V^\uparrow}^{(ij)}$ and $P_{(ij)_V}^{q^\uparrow}$ and
between $P_{s_V^\uparrow}^{(ij)}$ and $P_{(ij)_V}^{s^\uparrow}$ to be as follows:
\begin{eqnarray}
P_{q_V^\uparrow}^{(ij)}=1-P_{(ij)_V}^{q^\uparrow},~~
P_{s_V^\uparrow}^{(ij)}=1-P_{(ij)_V}^{s^\uparrow}.
\label{eqn:PSSA}
\end{eqnarray}

\subsection{Hadron-nucleus collision}
For a $hA$ collision, we assume that the projectile hadron successively
interacts with nucleons inside the nucleus $A$. 
 At each collision, the projectile hadron $h$ loses its momentum and
the rate of the momentum loss of $h$ is set to be $P(z)=0.25z^{0.25-1}$
from the data on the $A$-dependence of the spectra of $h$.
The probability of the incident hadron
colliding with $\nu$ nucleons, $P_{hA}(\nu)$, is calculated by using a Glauber-type multiple collision model.\cite{Glauber_Matthiae}
The number $\nu$ is determined from the distribution of nucleons in the nucleus and the cross
section of the incident hadron with a nucleon $\sigma_{hN}$. 
The nucleon number density of a 
nucleus with a mass number $A$ is assumed to be given by 
$\rho(r)=\rho_0/(1+\exp(\frac{r-r_A}{d}))$, 
where $\rho_0$ is the normalization factor.
We choose $r_A=1.19A^\frac{1}{3}-1.61A^{-\frac{1}{3}}$fm and $d=0.54$fm as used in Ref.\cite{qdqA}. 

\begin{figure}
\begin{center}
\resizebox{0.5\textwidth}{!}{%
  \includegraphics{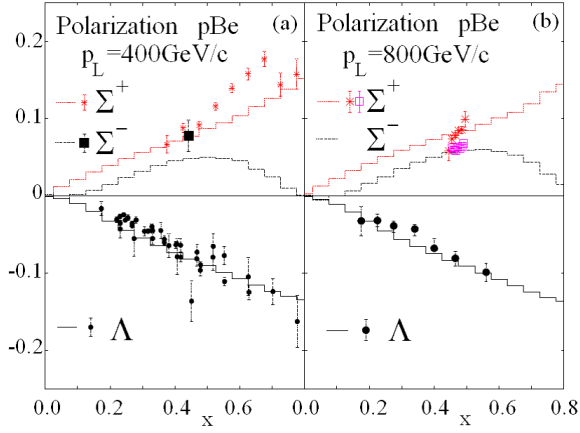}
}
\end{center}
\caption{The $x$ dependence of the hyperon polarizations in $pBe$ collisions at (a) 400 GeV/c and (b) 800 GeV/c
for $0.96$ GeVc$ < p_T $. Data are taken from Refs.4,7,8 and Refs.5,9.} 
\label{fgr:pBe400_800x}       
\end{figure}

\section{Comparison with the data} 
\label{sec:3}
\subsection{Setting of parameters in pBe collision}
\label{sec:3.1}
In the present analysis,
we assume that there is an intrinsic diquark-antidiquark state in the incident baryon with probability $c_1^2$. 
The intrinsic diquark recombines with the valence quark and becomes a non-leading baryon, 
but instead the intrinsic antidiquark behaves as a valence constituent. 
The intrinsic antidiquark 
is assumed to be in the spin-1 state. 
The probability of the (anti)diquark produced through (\ref{eqn:dq2M}) being in the spin-1 state is 
denoted by $P_D^1$. Here, we assumed that the intrinsic valence antidiquark has a property
identical to that of the valence diquark, i.e., it tends to combine with the spin-down sea antiquark.
We consider only $u, d$, and $s$ flavors, and the probability of $s\bar{s}$ pair creation is chosen to be
$P_{s\bar{s}}=0.12$. 
For other parameters irrelevant to polarization, 
we use the values of those determined in the previous analyses: 
$P_{PM}=0.15, \epsilon=0.07$ and $\eta=0.25$.\cite{tnki}

From the data on hyperon polarizations in $pBe$ collisions at $p_L=400$ GeV/c, we 
set the parameters as follows:
Since $\Lambda$ has a common diquark $[ud]_{00}$ with the incident proton, the 
spin dependent parameter $C_{[ud]}^s$ in (\ref{eqn:Pud00a}) is fixed from the data on $\Lambda$ polarization 
in $pBe$ collisions. 
Since the major part of $\Sigma^+$ production comes from the valence $\{uu\}^V_{11}$ diquark, 
the negative value of $C_{\{uu\}}^s$ leads to the positive polarization of $\Sigma^+$ in $pBe$ collisions. 
To explain the $\Sigma^+$ polarization, we have to choose a small value of $C_{\{uu\}}^s$. 
We set the asymmetry parameters as
\begin{eqnarray}
&C_{[ud]}^s&=C_{[ud]}^q=C_{[qs]}^s=-0.4, \nonumber \\
&C_{[qs]}^q&=-1.0, \\
&C_{\{qq\}}^s&=C_{\{qq\}}^q=C_{\{qs\}}^s=C_{\{ss\}}^q=C_{\{ss\}}^s=-1.0, \nonumber \\
&C_{\{qs\}}^q&=-0.2, 
\label{eqn:C_diquark}
\end{eqnarray}
where $q$ denotes $u$ and $d$ quarks. 
In order to explain the positive $\Sigma^-$ polarization in $pBe$ collisions, 
we have to choose a large value of $P_D^1$ and assume that the valence diquark
in the incident octet baryon is mainly in the spin-0 state: 
\begin{eqnarray}
P_D^1=0.95,~\cos^2\theta=0.9.
\label{eqn:P_D1_00}
\end{eqnarray}
The main production of leading $\Sigma^-,~\Xi^0$ and $\Xi^-$ comes from the converted spin-1 diquark and 
the polarizations of directly produced $\Sigma^-,~\Xi^0$ and $\Xi^-$ are positive.
Since the branching ratio of $\Sigma^*\rightarrow\Sigma\pi$ is 12$\%$, 
the effect of $\Sigma^*$ decay on $\Sigma^-$ polarization is small. 
On the other hand, 
$\Xi^0$ and $\Xi^-$ polarizations are affected by the 100$\%$ decay of 
$\Xi^*\rightarrow\Xi\pi$, leading to negative $\Xi^0$ and $\Xi^-$ polarizations. 
Here we note that the converted diquark through (\ref{eqn:dq2M}) from a valence diquark behaves like a valence diquark 
and produce a leading baryon.
From the data on $\bar{\Sigma}^-$ polarization, we set the parameter for the probability of the intrinsic 
diquark-antidiquark state in the incident baryon as 
\begin{eqnarray}
 c_1^2=0.07.
\label{eqn:P_intrnsc}
\end{eqnarray}

\begin{figure}
\begin{center}
\resizebox{0.25 \textwidth}{!}{%
  \includegraphics{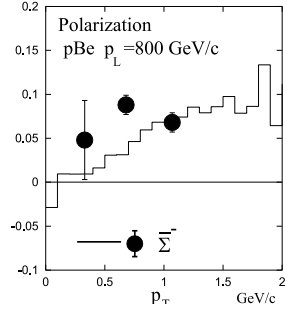}
}
\end{center}
\caption{The $p_T$ dependence of the $\bar{\Sigma}^-$ polarization in $pBe$ collisions at $p_L=$800 GeV/c 
for $ 0.47 < x < 0.53$. 
Data are taken from Ref.13.} 
\label{fgr:pBe800Sig_bar}       
\end{figure}

The $x$ dependence of the $\Lambda$ and $\Sigma^\pm$ polarizations at $p_L$ = 400\cite{Lundberg,Wilkinson,Deck} 
and 800\cite{Ramberg94,Morelos95} GeV/c
in $pBe$ collisions is shown in Fig.\ref{fgr:pBe400_800x}. The $p_T$ dependence of
the $\bar{\Sigma}^-$ polarization at $p_L$ = 800 GeV/c in $pCu$ collisions is shown in Fig.\ref{fgr:pBe800Sig_bar}\cite{Morelos93}.
If $c_1^2=0$, the $\bar\Sigma^-$ polarization is zero. 
Thus, the second term in (1) plays an important role on the polarization of the anti-hyperon in our model.

\begin{figure}
\begin{center}
\resizebox{0.5 \textwidth}{!}{%
  \includegraphics{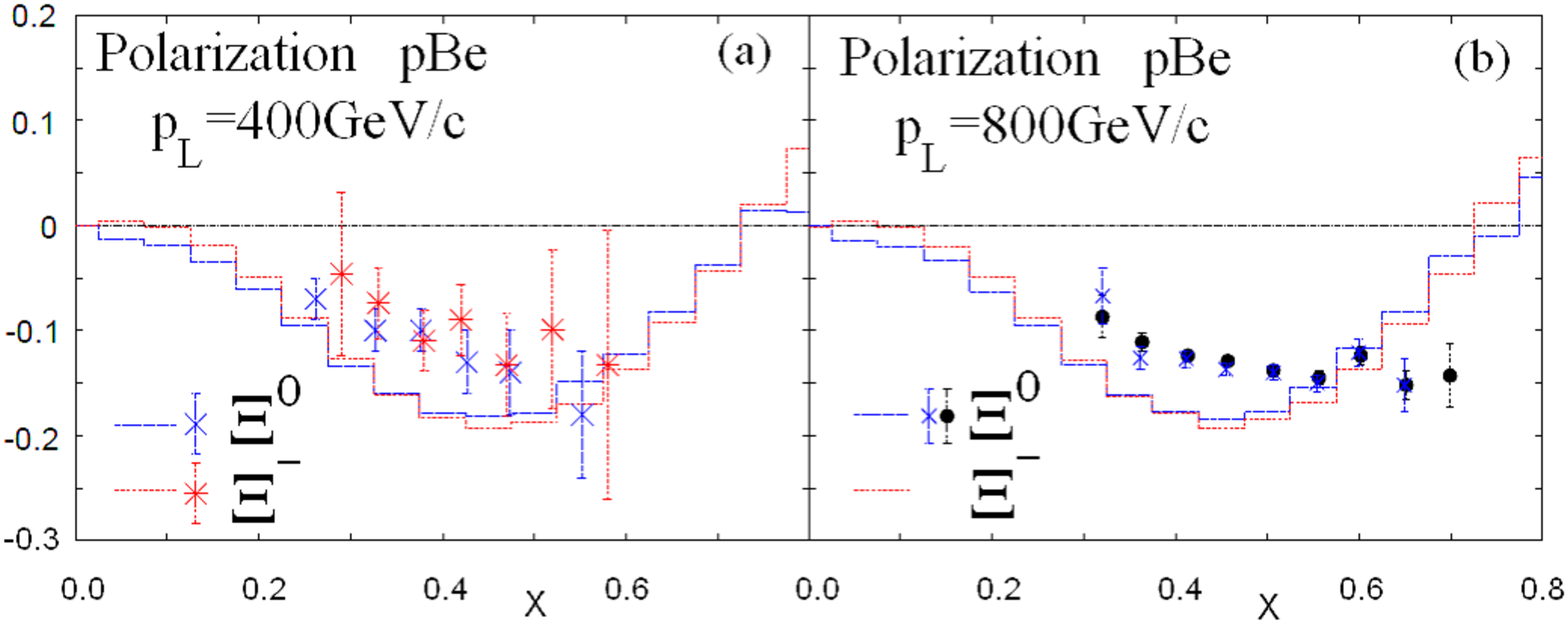}
}
\end{center}
\caption{The $x$ dependence of $\Xi$ polarizations in 
in $pBe$ collisions at (a) 400 GeV/c and (b) 800 GeV/c. Data are taken from Refs.10,11 and Ref.12.} 
\label{fgr:pBe400_800Xi}       
\end{figure}
\begin{figure}
\begin{center}
\resizebox{0.5 \textwidth}{!}{%
  \includegraphics{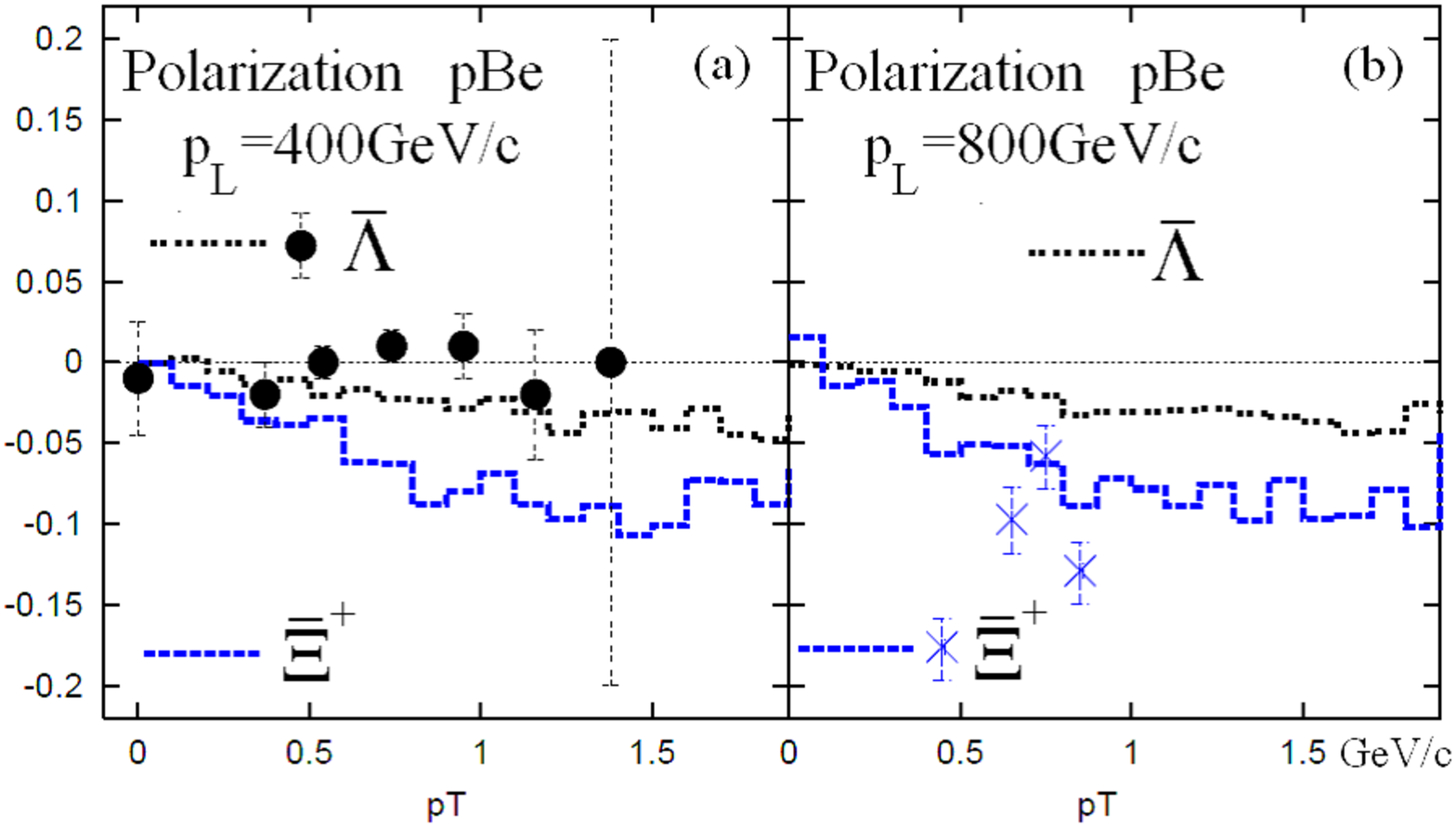}
}
\end{center}
\caption{
The $p_T$ dependence of anti-hyperon polarizations in $pBe$ collisions at
$p_L=400$ and 800 GeV/c. Data on $\bar{\Lambda}$ and $\bar{\Xi}^+$ are taken from Ref.6 and Ref.14.} 
\label{fgr:pBe_anti-LamXi}       
\end{figure}

\begin{figure}
\begin{center}
\resizebox{0.7 \textwidth}{!}{%
  \includegraphics{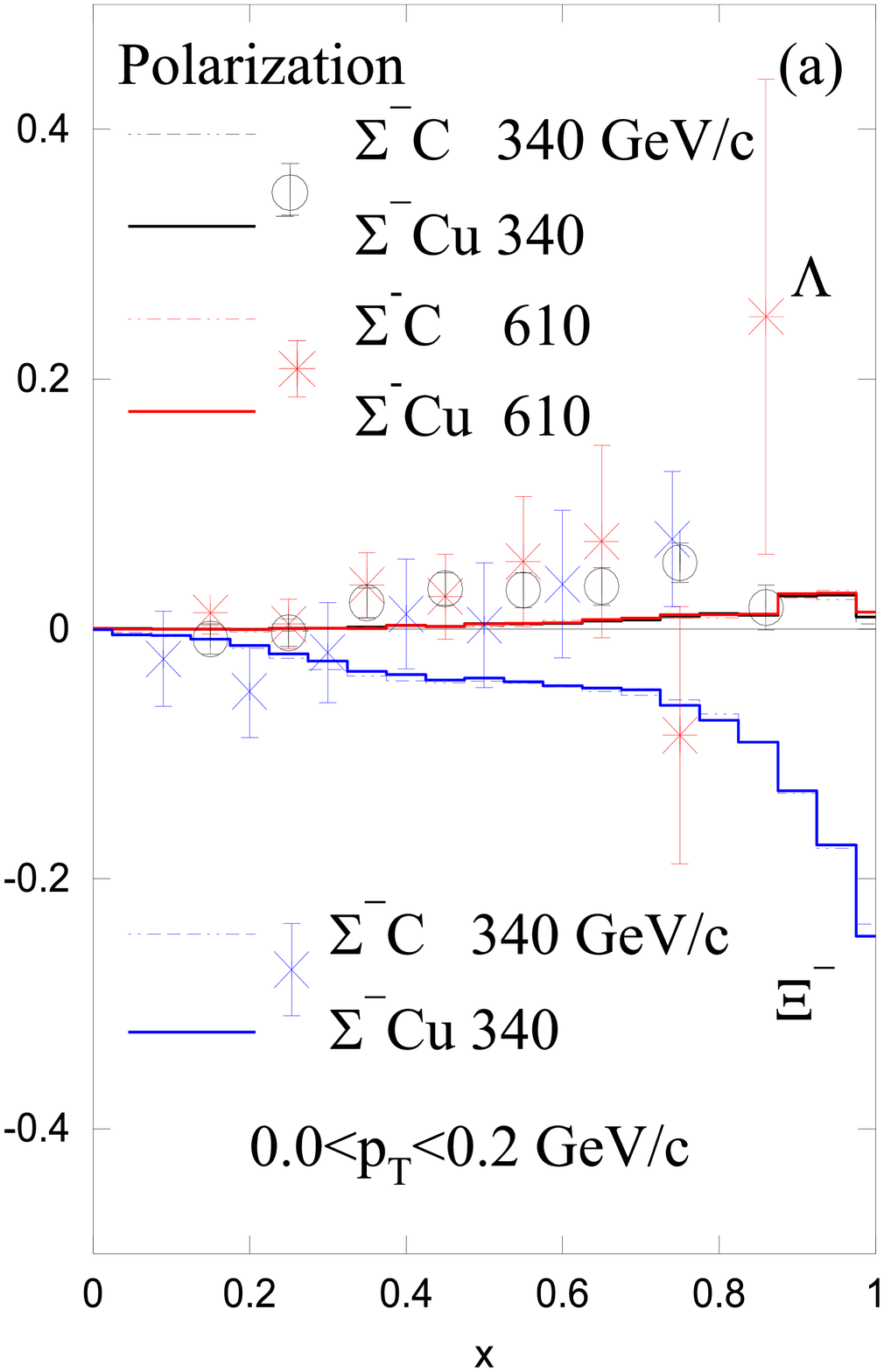}
  \includegraphics{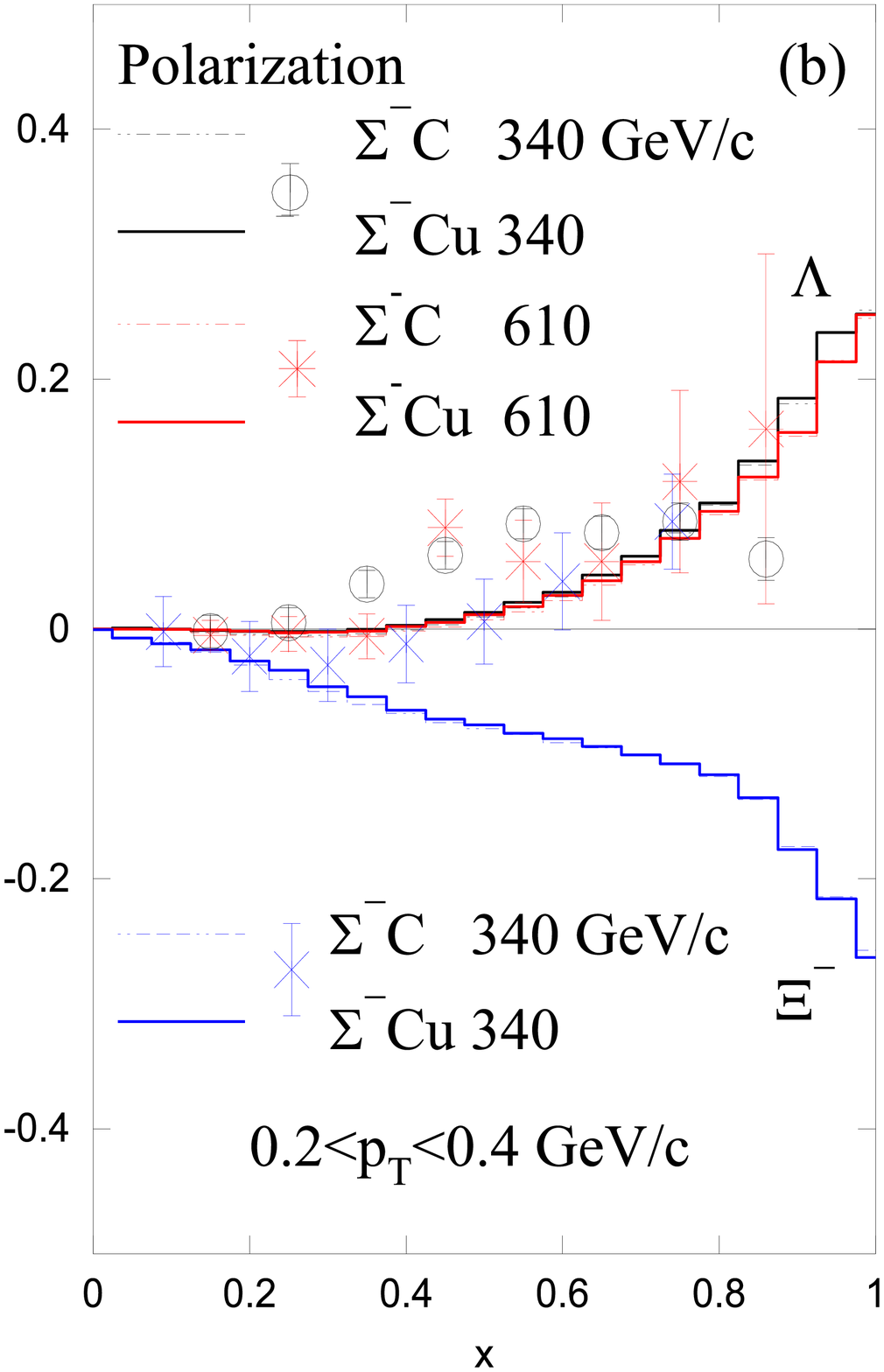}
  \includegraphics{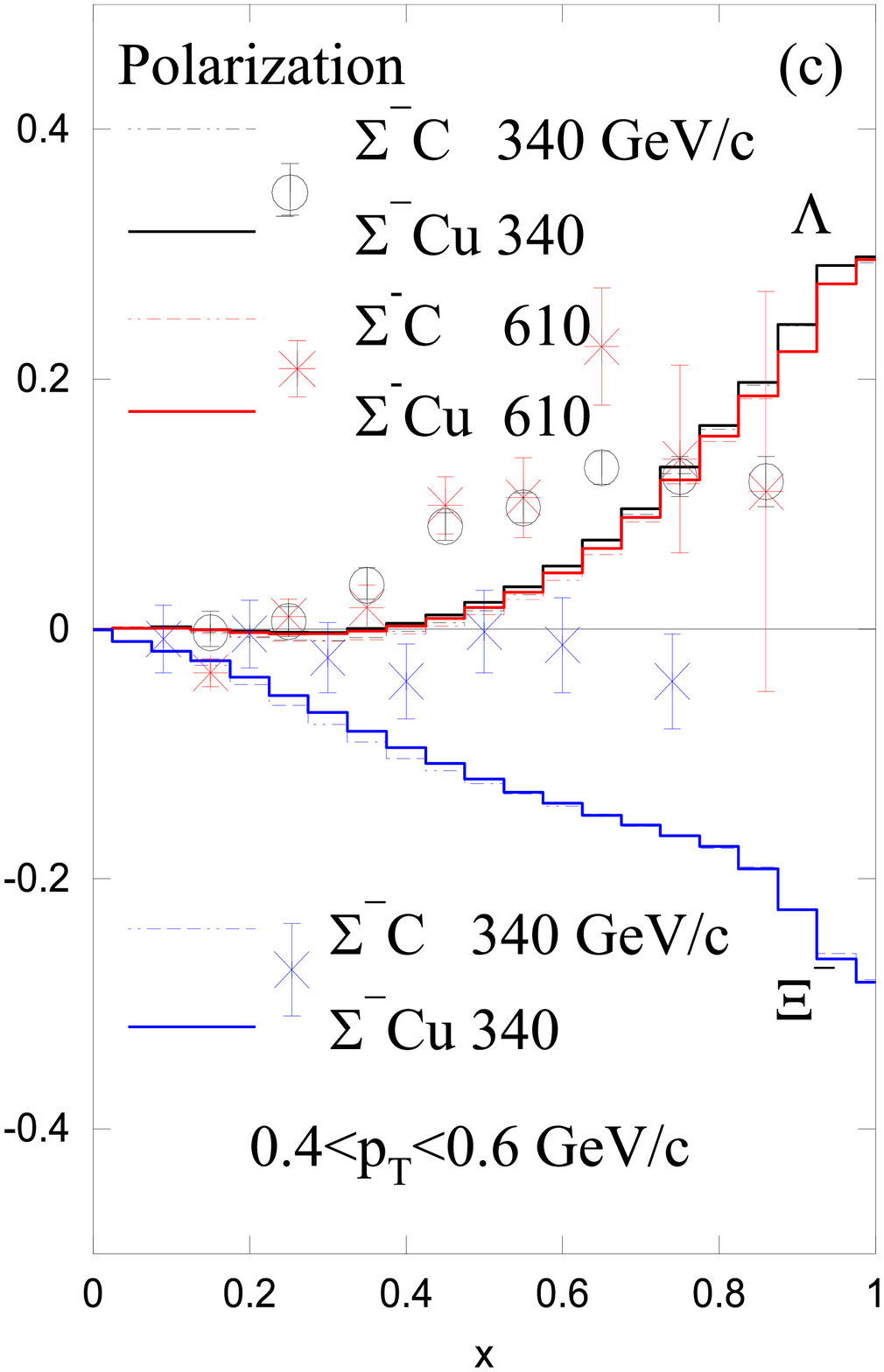}
}
\resizebox{0.7 \textwidth}{!}{%
  \includegraphics{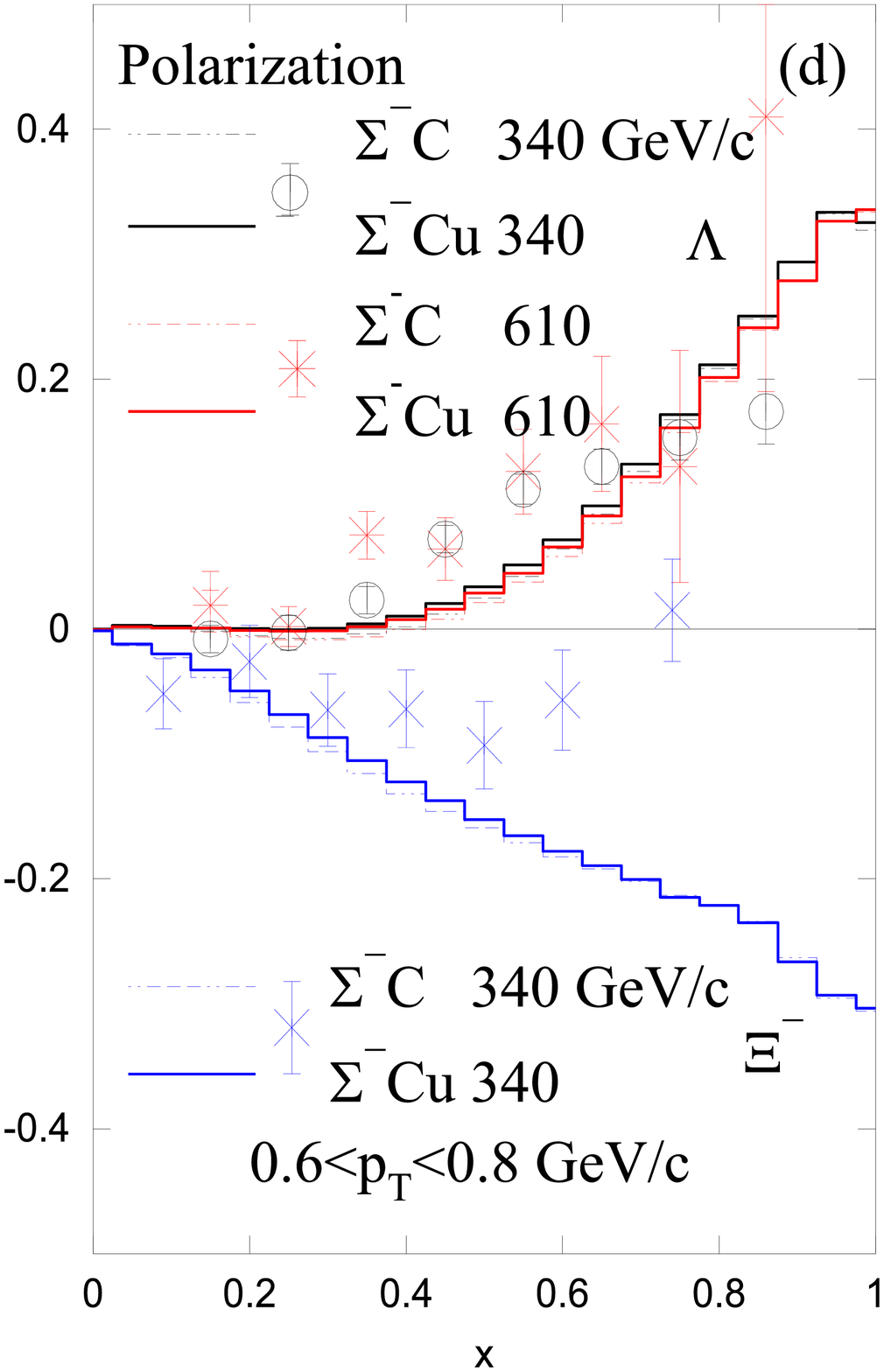}
  \includegraphics{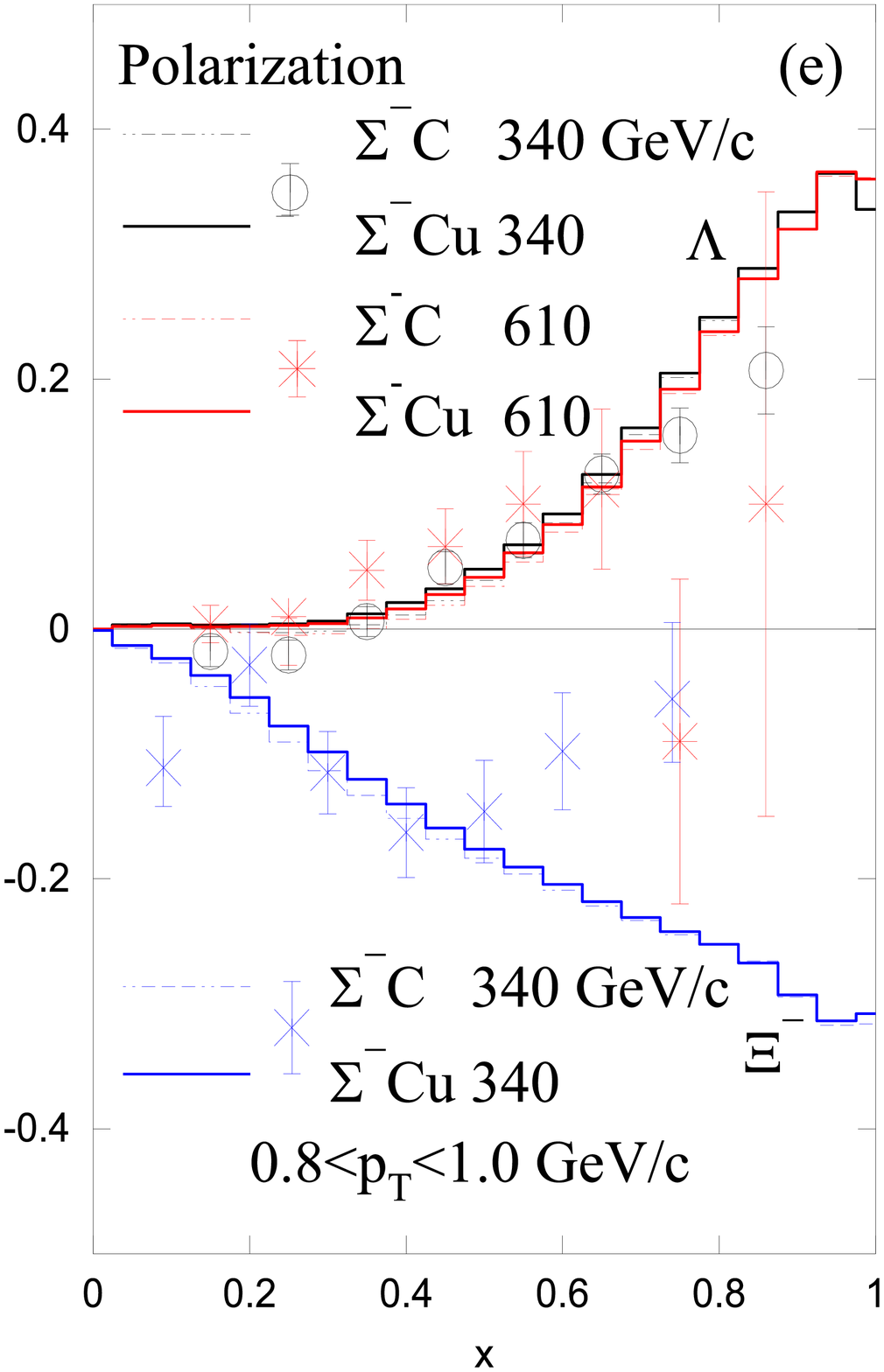}
  \includegraphics{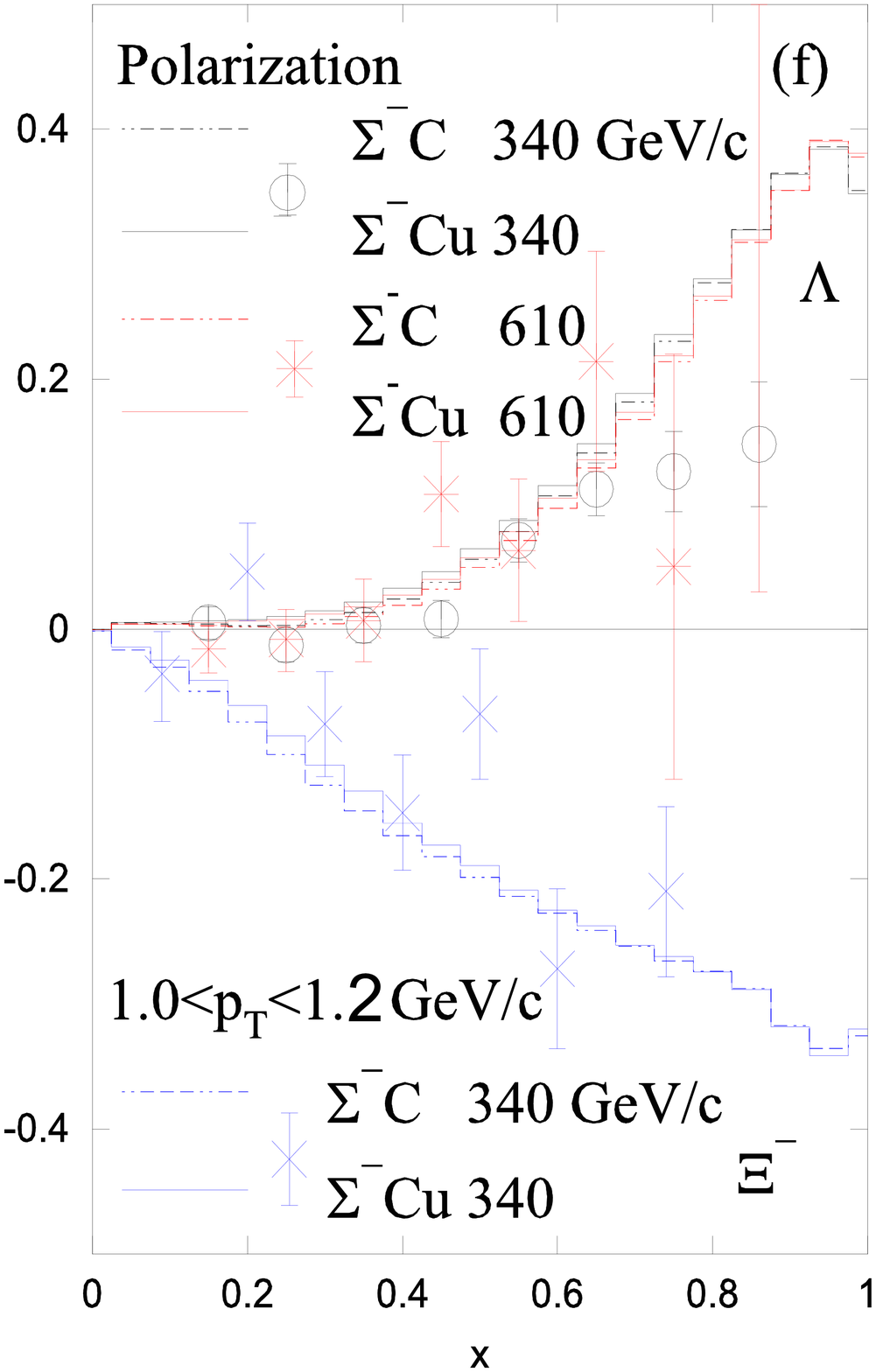}
}
\end{center}
\caption{
The $x$ dependence of $\Lambda$ and $\Xi^-$ polarizations in 
$\Sigma^-C$ and $\Sigma^-Cu$ collisions at $p_L=340$ and 610 GeV/c in the regions (a) $0<p_T<0.2$, (b) $0.2<p_T<0.4$, 
(c) $ 0.4 < p_T < 0.6$, (d) $0.6<p_T<0.8$, (e) $0.8 < p_T < 1.0$ and (f) $1.0<p_T<1.2$ GeV/c. 
Data are taken from Refs.48,49,50.}
\label{fgr:Sig-Cu340_LamXi}       
\end{figure}

\subsection{Model predictions on other processes}
In this subsection, we present the results of our model for polarizations of 
other hyperons in  unpolarized $pA$ and $\Sigma^-A$ collisions. 
In Fig.\ref{fgr:pBe400_800Xi}, the $x$ dependence of $\Xi$ polarization in $pBe$ collisions is compared with the data 
at $p_L=400$\cite{Heller83,Rameika86} 
and $800$\cite{Duryea91} GeV/c. 
In Fig.\ref{fgr:pBe_anti-LamXi}, the results of the polarizations of $\bar{\Lambda}$ and $\bar{\Xi}^+$ in $pBe$ collisions 
are compared with experimental data at $p_L=400$\cite{Heller78} and 800\cite{Ho} GeV/c, respectively. 
By choosing the weight of the intrinsic antidiquark to be $c_1^2=0.07$,
we obtain good agreement with experimental data. 

Fig.\ref{fgr:Sig-Cu340_LamXi} shows the $x$ dependence of $\Lambda$ and $\Xi^-$ polarizations 
in $\Sigma^-C$ and $\Sigma^-Cu$ collisions at $p_L=340$\cite{Adamovich_Lambda,Adamovich_Xi-} and 610 GeV/c\cite{Sanchez-Lopez}. 
Since the incident $\Sigma^-$ is mainly made out of $[ds]_{00}^V$ and $d_V$, 
the negative value of $C_{[qs]}^q$ in (\ref{eqn:C_diquark}) leads to negative $\Lambda,\Sigma^0$ and $\Xi^-$ polarizations. 
The leading decuplet baryon production is much suppressed 
and the resonance effect of decuplet baryon on hyperon polarization is small. 
However, the leading $\Sigma^0$ production is three times larger than the leading $\Lambda$ production 
in $\Sigma^-$ projectile from SU(6) wave function. 
Taking into account the 100 $\%$ branching
ratio for $\Sigma^0 \rightarrow \Lambda\gamma$ and the spin conservation, 
one expects positive $\Lambda$ polarization in the $\Sigma^-Cu$ collision. 
In the proton projectile, there is no leading $\Sigma^0$ production and the $\Lambda$ polarization remains negative.
The calculated results for $\Xi^-$ polarization at large $p_T$ show negative values, in agreement with 
the experimental data\cite{Adamovich_Xi-}. 

The $p_T$ dependence of $\Sigma^+$ polarization in $\Sigma^-C$ and $\Sigma^-Cu$ collisions at $p_L$ = 330 GeV/c 
is shown in Fig.8(a)\cite{Adamovich_Lambdabar_Sig+}. 
Although the main part of the converted diquark through (\ref{eqn:dq2M}) is in the spin-1 state, 
the presence of spin-0 diquarks $[us]$ is inferred from
the negative $\Sigma^+$ polarization. 
In Fig.8(b), the results for anti-hyperons in $\Sigma^-C$ and $\Sigma^-Cu$ collisions at $p_L=330$ GeV/c are shown\cite{Adamovich_Lambdabar_Sig+}. 
The polarization of $\bar{\Lambda}$ is small, similar to the case of the $pBe$ collision.

\begin{figure}
\begin{center}
\resizebox{0.5 \textwidth}{!}{%
  \includegraphics{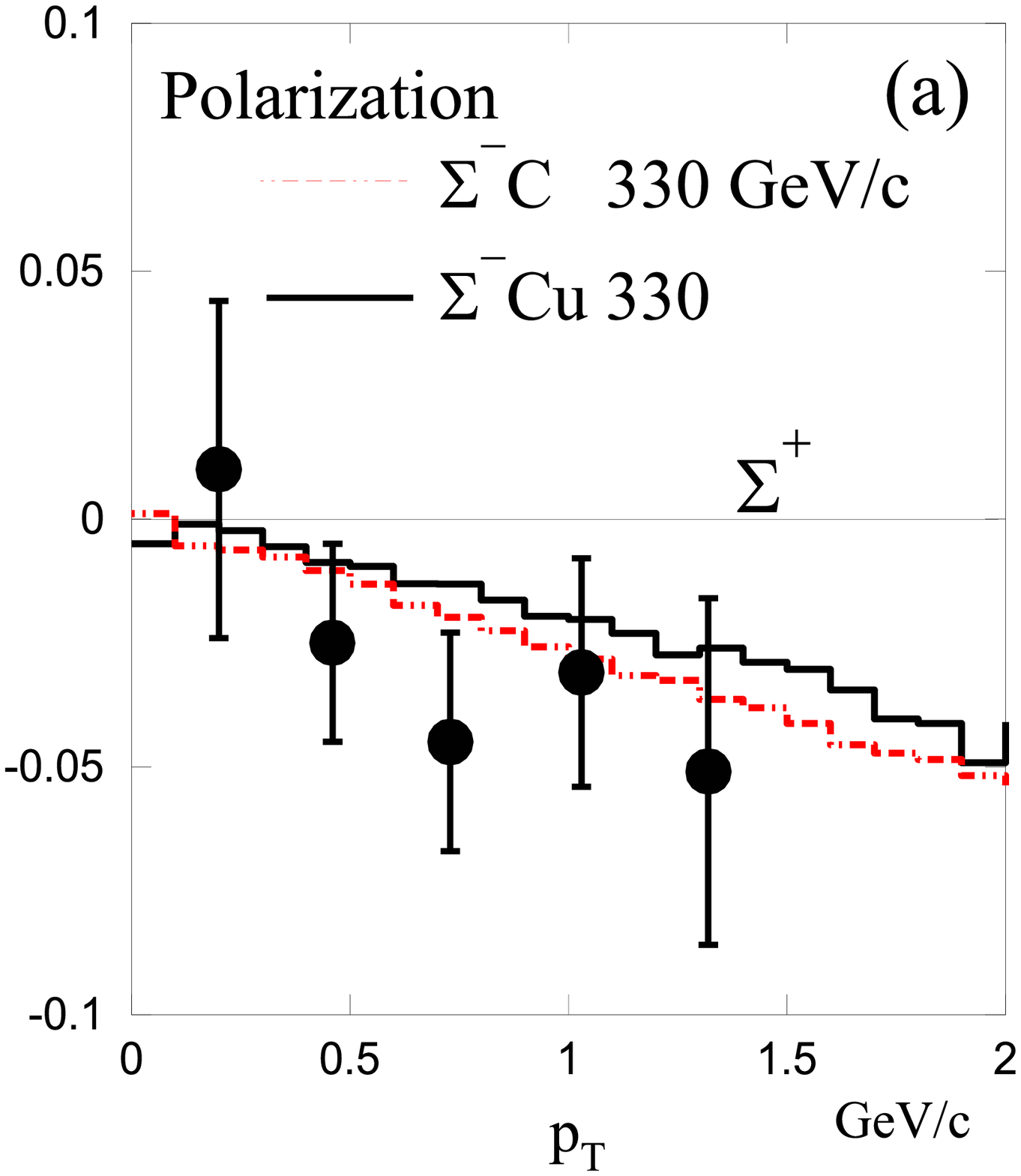}
  \includegraphics{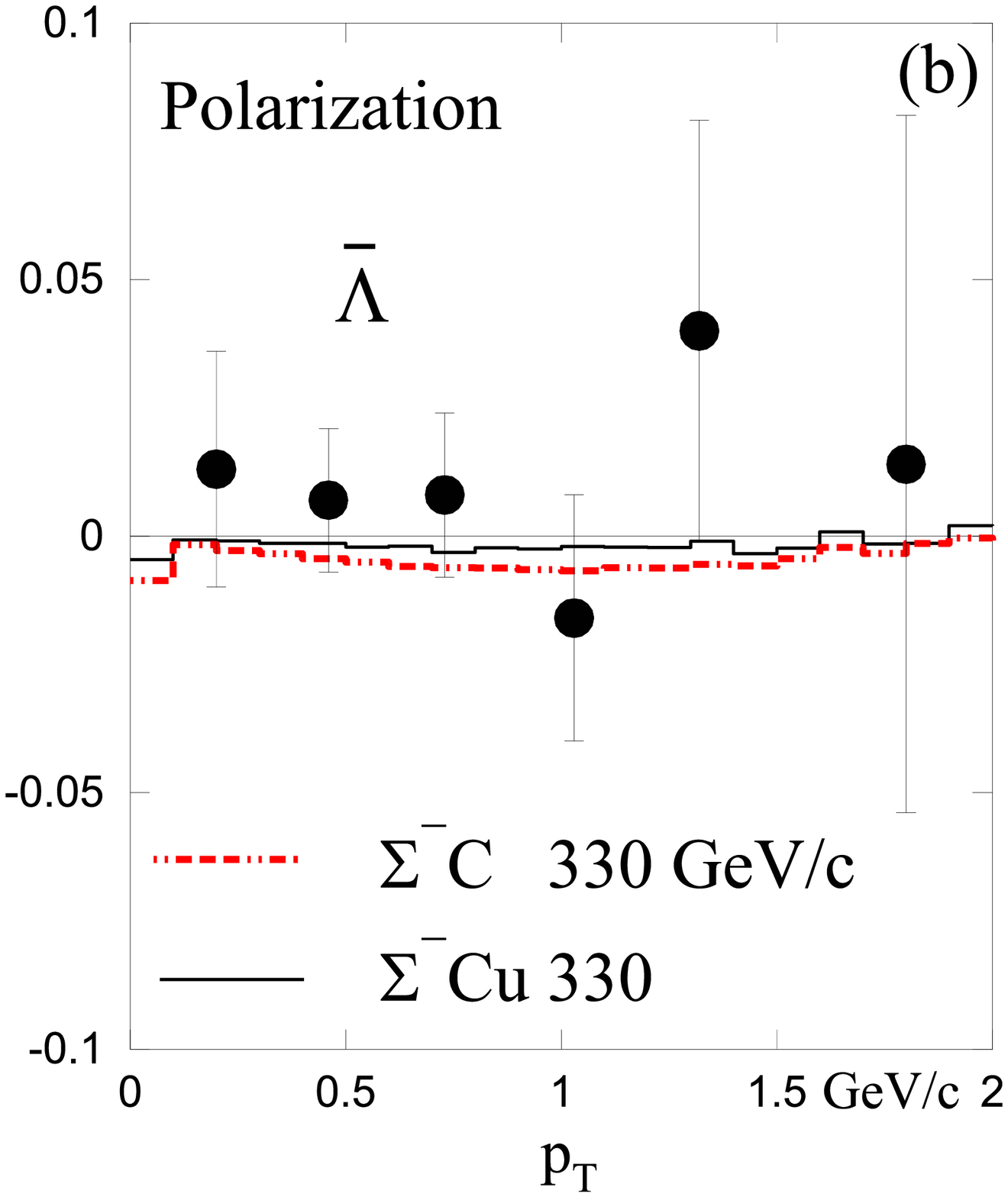}
}
\end{center}
\caption{The $p_T$ dependence of the polarizations of (a) 
$\Sigma^+$ and (b) 
$\bar{\Lambda}$ polarizations 
in $\Sigma^-C$ collisions at $p_L=330$ GeV/c.
Data are taken from Ref.51. 
}
\label{fgr:Sig-Cu_Lambar+}      
\end{figure}

\subsection{Results from analysis}
\label{sec:3.3}
We show the $x$ distribution functions $n(x)$ of constituents 
in the incident baryons ($p$ and $\Sigma^-$) in Fig.9 as results from this analysis. 
For comparison, we show the CTEQ3 input parton distribution functions at $Q_0=1.6$GeV of a global QCD analysis 
for various hard scattering processes\cite{Lai}. 
In our model, it is assumed that incident baryons are mainly made out of 
valence spin-0 diquark and valence quark. 
Consequently, the magnitude of valence quarks are small as compared with those of CTEQ3 distributions.
Since main part of sea (anti)quarks are produced in the cascade processes (\ref{eqn:q2B})-(\ref{eqn:dq2M}), the magnitude of 
sea (ant-)quarks in the incident baryons are much suppressed as shown in Fig.9(b). 
The $x$-shape behavior of CTEQ3 distribution functions of valence and sea quarks in hard interaction 
are different from our results in soft interaction. 

\begin{figure}
\begin{center}
\resizebox{0.6 \textwidth}{!}{%
  \includegraphics{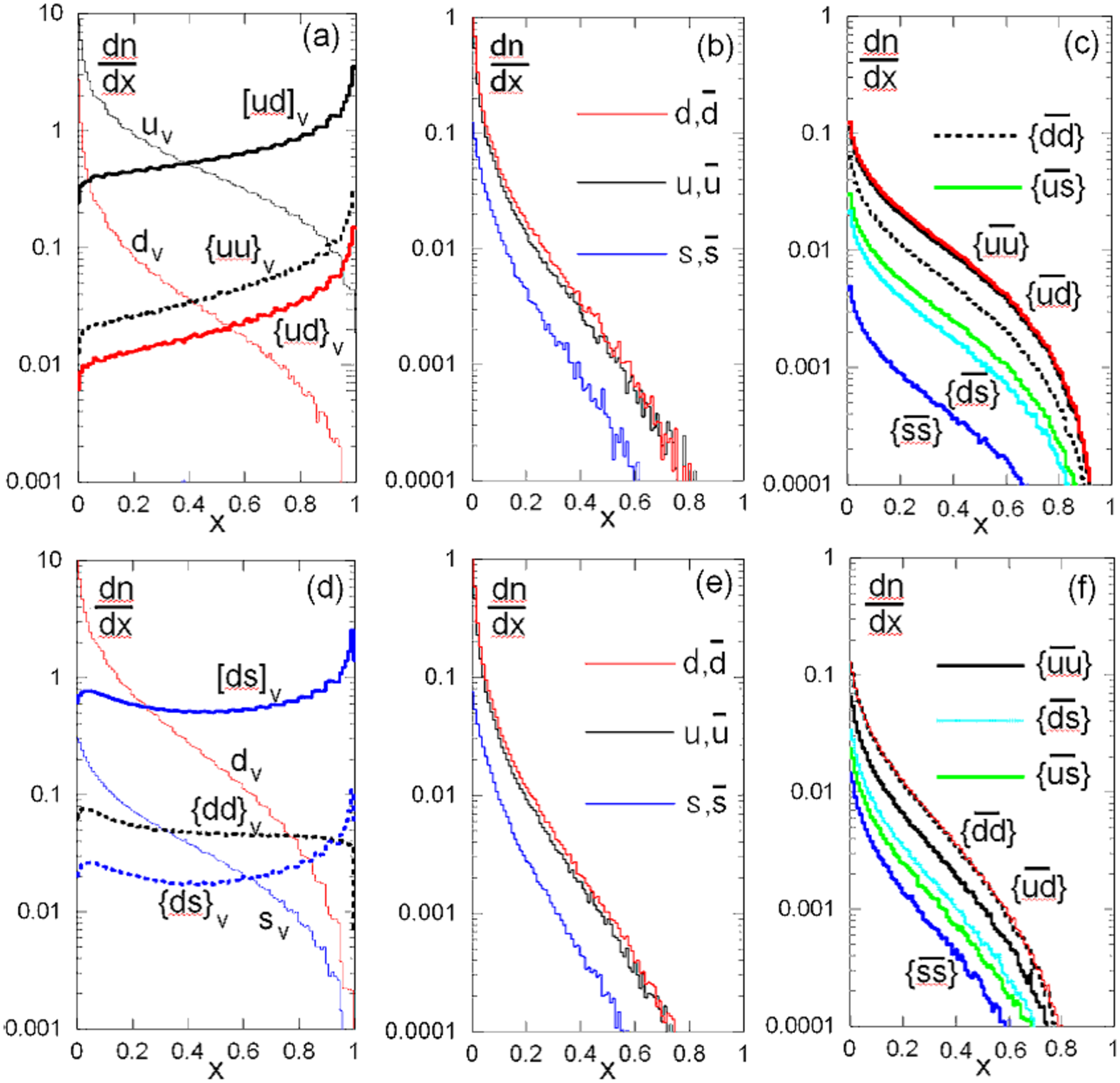}
}
\end{center}
\caption{
 The $x$ distributions of constituents in the incident baryons. (a) valence diquarks, (b) (anti)quarks and (c) anti-diquarks in 
$pBe$ collisins at $p_L=800$ GeV/c. The lower half is the same for $\Sigma^-Cu$ collisions 
at $p_L=330$ GeV/c. 
}
\label{fgr:valence}       
\end{figure}

\section{Conclusion and discussion}
\label{sec:4}
In our model, it is assumed that 
(i) the quantization axis is characterized by the leading baryon of the most massive cascade chain, 
(ii) the incident valence diquark tends to pick up a spin-down sea quark (or conversely the spin-up incident 
valence quark tends to turn to the left and combines 
with a sea diquark to form a leading baryon), and 
(iii) the incident baryon contains an intrinsic antidiquark that produces a leading anti-baryon. 
Our model with the parameters in subsection 3.1
explains the hyperon and anti-hyperon polarizations both in $pA$ and $\Sigma^-A$ collisions well. 
From our analysis, we expect that 
the incident spin-1/2 baryon is mainly composed 
of a spin-0 valence diquark and a valence quark, but contains about 10$\%$ of spin-1 valence diquarks. Further, it appears that 
there is an intrinsic antidiquark with a probability of about 7$\%$. 

The complex hyperon polarizations may be explained by considering the leading particle effects of valence constituents
and by taking into account the contributions of the decay products from resonances.
Polarizations of hyperons having a common diquark with the projectile,
such as $\Sigma^+, \Sigma^0$ and $\Lambda$, in unpolarized $pA$ collisions hardly depend on the center-of-mass energy $\sqrt{s}$,
and those of hyperons having only one common quark with the projectile, such as $\Sigma^-, \Xi^0$ and $\Xi^-$ 
in $pA$ collisions, depend on $\sqrt{s}$. The target mass number dependence of
polarizations of hyperons is low.
The energy dependence of hyperon polarization in $\Sigma^-A$ collisions is small due to the strangeness
of the incident valence diquark.

Our approach is applicable to small $p_T$ regions ($p_T < 2$GeV/c), while the    
approaches based on QCD factorization are mainly applicable to large $p_T$ regions ($p_T > 2$GeV/c). 
Thus, there is no direct correspondence between these approaches. However, it is an important problem to research a 
relation of our model to QCD factorization model. We will investigate this problem as one of future works.

%
%
%

\end{document}